\documentclass[aps,prd,11pt,notitlepage,nofootinbib,superscriptaddress,showkeys,showpacs]{revtex4-1}
\usepackage{amstext,amsmath,amssymb,amsfonts,bbm}
\usepackage[latin1]{inputenc}
\usepackage{graphicx}
\usepackage{hyperref}
\usepackage{epstopdf}
\usepackage{amsthm}
\usepackage{pstricks}

\usepackage{color}

\def\beq{\begin{equation}}
\def\be{\begin{equation}}
\def\ee{\end{equation}}
\def\bes{\begin{eqnarray}}
\def\ees{\end{eqnarray}}

\DeclareMathOperator{\Ad}{Ad}

\DeclareMathOperator{\id}{id}

\DeclareMathOperator{\SU}{SU}
\DeclareMathOperator{\SO}{SO}

\newcommand{\su}{\mathfrak{su}}

\newcommand{\unit}{\mathbbm{1}}

\def\f{\frac}

\def\mone{^{-1}}
\def\tl{\widetilde}

\def\what{\widehat}

\def\eps{\epsilon}

\def\C{{\mathbbm C}}
\def\N{{\mathbbm N}}

\def\R{{\mathbbm R}}

\def\calA{{\mathcal A}}

\def\calH{{\mathcal H}}

\theoremstyle{definition}
\theoremstyle{definition}
\theoremstyle{definition}
\theoremstyle{definition}
\theoremstyle{definition}
\theoremstyle{definition}

\begin{document}

\title{\Large \bf The Hamiltonian constraint in 3d Riemannian loop quantum gravity}

\author{{\bf Valentin Bonzom}}\email{vbonzom@perimeterinstitute.ca}
\author{{\bf Laurent Freidel}}\email{lfreidel@perimeterinstitute.ca}

\affiliation{Perimeter Institute for Theoretical Physics, 31 Caroline St. N, ON N2L 2Y5, Waterloo, Canada}

\date{\small\today}

\begin{abstract}\noindent
We discretize the Hamiltonian scalar constraint of three-dimensional Riemannian gravity on a graph of the loop quantum gravity phase space. This Hamiltonian has a clear interpretation in terms of discrete geometries: it computes the extrinsic curvature from dihedral angles. The Wheeler-DeWitt equation takes the form of difference equations, which are actually recursion relations satisfied by Wigner symbols. On the boundary of a tetrahedron, the Hamiltonian generates the exact recursion relation on the 6j-symbol which comes from the Biedenharn-Elliott (pentagon) identity. This fills the gap between the canonical quantization and the symmetries of the Ponzano-Regge state-sum model for 3d gravity.

\end{abstract}

\maketitle


\section{Introduction}

Three-dimensional Riemannian gravity is now a quite well understood theory, much easier than the four-dimensional case since it does not carry local degrees of freedom. In spite of its apparent simplicity, it has been a good testing ground to build new methods aiming at a background independent quantization of gravity. In particular, the quantization programs of loop quantum gravity and spin foams have been successfully applied, on the kinematical aspects of quantum geometry \cite{rovelli-PRTVO, livine-3dlength} as well as on its dynamics \cite{thiemann-3d, ooguri-3d, freidel-louapre-PR1, freidel-louapre-PR2,noui-perez-ps3d, meusburger-combinatorial, noui-perez-particle}.

Still, there remain two important questions in 3d gravity, one in the canonical loop quantum gravity (LQG) approach, and the second relating to the spin foam formalism. Here, we will propose a common solution to both, showing they are two aspects of the same problem. To say things briefly, a direct and natural quantization of the scalar constraint in 3d gravity, which would reproduce known results, is still missing (see \cite{thiemann-3d} though, and the discussion below), while in the Ponzano-Regge spin foam approach, recurrence relations on the basic amplitudes, coming from the topological symmetry of the model, still need to be generated from the canonical formalism.

Through the paper, we will consider pure three-dimensional gravity as a Chern-Simons-BF theory. This is a topological gauge theory whose reduced phase space is the cotangent bundle to the moduli space of flat $G$ connections (here with $G=\SU(2)$). Such theories (like two-dimensional Yang-Mills, three-dimensional Chern-Simons, and BF in higher dimensions) display some intriguing relations with the representation theory of the local symmetry group when one wants to quantize before imposing the symmetries. Typically, one can write their partition functions as state-sums based on the 6j-symbol (or higher Wigner symbols, or $q$-deformed versions). Then the symmetry which makes the model topological is described by algebraic relations on the 6j-symbols, such as the Biedenharn-Elliott (BE), pentagon identity.

Usually, the canonical quantization in the Hamiltonian framework is a preferred setting to study the symmetries. Hence one should be able to derive such algebraic characterizations from the generators of the symmetries in the Hamiltonian formalism. As for three dimensional gravity itself, the framework of LQG seems very suitable since the boundary data of the corresponding state-sum model (the Ponzano-Regge model \cite{freidel-louapre-PR1}) are the kinematical quantum numbers of LQG (on a graph dual to a triangulation). This is the reason why a specific project of studying the symmetries of simple models from the LQG point of view was initiated in \cite{recurrence-paper} using recursions on recoupling Wigner symbols and generalizations. The present paper is a natural follow-up which goes further in the 3d gravity case. Indeed, in spite of the well-established fact that the moduli space of flat $\SU(2)$ connections can be reached from $\SU(2)$ recoupling theory \cite{ooguri-3d, barrett-crane-wdw, freidel-louapre-PR1, noui-perez-ps3d} (and more recently \cite{twisted-bubbles}), there has been no derivation of the key recursion on the 6j-symbol from Hamiltonian generators in LQG (but see an interpretation in \cite{barrett-crane-wdw}),
\be \label{rec6j}
A_{+1}(j)\,\begin{Bmatrix} j_1+1 &j_2 &j_3 \\ j_4 &j_5 &j_6 \end{Bmatrix} +
A_{0}(j)\,\begin{Bmatrix} j_1 &j_2 &j_3 \\ j_4 &j_5 &j_6 \end{Bmatrix} +
A_{-1}(j)\,\begin{Bmatrix} j_1-1 &j_2 &j_3 \\ j_4 &j_5 &j_6 \end{Bmatrix} = 0\;.
\ee
This is a second order difference equation (whose coefficients will be given in the main text) which can be derived from the BE identity and in turn fully determines the 6j-symbol \cite{varshalovich-book, schulten-gordon1}.

The above discussion should be put in parallel with the fact that we do not know the Wheeler-DeWitt equation in 3d LQG,
\beq \label{wdw eq}
\widehat{H}\vert \psi\rangle=0\;,
\ee
i.e. the quantum equation for the Hamiltonian scalar constraint $H$ in the spin network basis. Spin network states in LQG are labeled by spins attached to links of embedded graphs. Hence, if the Hamiltonian is not graph changing (as expected in three dimensional gravity), the Wheeler-DeWitt should be a difference equation, with non-trivial coefficients. It should be possible to solve it so as to reproduce results known from other approaches, and make contact with the symmetries encoded in spin foams.  The recursion \eqref{rec6j} is obviously the natural candidate, when the 6j-symbol is the physical state, like on the 2-sphere triangulated with four triangles (details in the main text).

We consider in the present note a classical quantity $H_{v,f}$ on the phase space of LQG dual to a triangulation and turn it into an operator $\widehat{H}_{v,f}$ which enjoys the following nice properties.
\begin{itemize}
 \item $H_{v,f}$ can be understood as a discretization of the scalar Hamiltonian constraint and forms a first class system (we only checked that on a specific triangulation of the 2-sphere).
 \item It has a nice, natural interpretation in terms of discrete geometries and dihedral angles of flat simplices.
 \item $\what{H}_{v,f}$ acts on spin network functionals, and is labeled by a \emph{vertex} $v$ of a spin network graph and a \emph{cycle} containing it -- this cycle bounds the face $f$.
\end{itemize}
The main result is that $H_{v,f}$ is the missing link between the scalar constraint in LQG and the BE identity.
\begin{itemize}
 \item The equation $\widehat{H}_{v,f}\,\vert \psi\rangle=0$, when written in the spin network basis, produces difference equations of order 2.
 \item For trivalent faces, the difference equation is \eqref{rec6j} and hence solved uniquely by 6j-symbols, which project the curvature around the face to zero, as desired (for the solutions which we expect more generally, see \cite{ooguri-3d, witten-3d-ampli}, and correspondingly some expected recursions can be found in \cite{varshalovich-book, recurrence-paper}).
\end{itemize}

Even if we cannot fully solve the model with our new proposal for the Hamiltonian yet (because the case of faces of more than three links requires recursion relations on arbitrary Wigner symbols), it is worth comparing with the previous LQG quantization. In \cite{noui-perez-ps3d}, the authors use a simpler form of the flatness constraint which makes the quantization easier, but is far from being applicable to four dimensional gravity. Nevertheless, the contact with $\SU(2)$ recoupling is natural is this case (see also \cite{ooguri-3d} which is very close), though not complete with respect to our proposal. A completely different proposal was made in \cite{thiemann-3d}, to quantize directly the scalar constraint. The author made use of what is now known as the Thiemann's trick, whose advantage is that it can be straightforwardly extended to four dimensions, and a major success is certainly that it leads to an anomaly-free algebra. However diffeomorphisms of the canonical surface and the scalar constraint are not treated on equal footing, in contrast with what happens in state-sum models. We will argue that our discretization of the scalar constraint actually and surprisingly contain all the constraints on the graph. This is certainly the deep reason why the quantization of \cite{thiemann-3d} does not exhibit any clear link with the recoupling theory and the BE identity.

In section \ref{sec:review}, we review some elementary facts about 3d Riemannian gravity, its loop quantization, and the physical state on the 2-sphere. Then, in section \ref{sec:class}, we present our proposal for the Hamiltonian operator on spin networks, describe its classical geometric interpretation and show it is first class (on a specific triangulation). In section \ref{sec:quantum}, we proceed to the quantization, and prove the relation to the flatness constraint on trivalent faces. Finally, in section \ref{sec:spin2}, we discuss higher order difference equations, coming from lifting our Hamiltonian to representations of higher spins.

\section{Review of 3d Loop Quantum Gravity} \label{sec:review}
\subsection{Canonical analysis of 3d gravity}

Three-dimensional Riemannian gravity can be formulated with a connection $A$ on a principal $\SU(2)$-bundle $P$ over a three-manifold $M$, and a cotriad $e$, that is an $\Ad(P)$-valued 1-form. Locally, the connection is seen as a Lie algebra $\su(2)$-valued 1-form over $M$, written $A^i$ for $i=1,2,3$, with a specific rule for gauge transformations. Similarly, $e$ is locally a 1-form over $M$ transforming through the adjoint representation. If $g$ is map from $M$ to $SU(2)$, the corresponding gauge transformation reads $A \mapsto \Ad(g) A + g\,dg\mone$ and $B \mapsto \Ad(g) B$. The adjoint action is in matrix form: $\Ad(g) X = gX g\mone$. To write the action, note that $\su(2)$ is equipped with a non-degenerate bilinear invariant form, which enables to raise and lower indices via the Kronecker delta $\delta_{ij}$. The curvature of $A$ is the $\su(2)$-valued 2-form $F(A)=dA+\f{1}{2}[A,A]$, and the action is
\beq
S(e,A) = \frac{1}{4\pi G} \int_M e^i\wedge F(A)^j\ \delta_{ij}\;.
\ee
In the following, we will set $4\pi G=1$.

The equations of motion are the torsion free condition, $d_A\,e=de+[A,e]=0$, and the flatness condition, $F=0$. It can be shown that all solutions are locally equivalent to the trivial one, $A=0$ and $e=0$. This fact is due to an additional symmetry of the action,
\beq \label{translation sym}
A \mapsto A\;, \qquad e \mapsto e + d_A\,\eta\;,
\ee
for any function $\eta$. This symmetry makes possible to locally gauge the cotriad to zero.

The Hamiltonian analysis reproduces the covariant picture, while preparing the road to quantization \`a la Dirac. To see how this happens, assume $M$ is of the form $\Sigma\times \R$, where $\Sigma$ is a smooth two-dimensional manifold of arbitrary topology. Choose arbitrary coordinates $x^a=(x^1,x^2)$ over the canonical surface $\Sigma$, together with a 'time' coordinate $t$ on $\R$. Following this splitting, the action admits the decomposition
\beq
S(e,A) = \int dt\int d^2x\ \left[\epsilon^{ab}\,\delta_{ij}\,\left(e_a^i\,\partial_t A_b^j\right) \,+\,\delta_{ij}A_t^j\,\left(\epsilon^{ab}\,D_a^{\phantom{i}} e_b^i\right) \,+\, \delta_{ij}e_t^j\,\left(\epsilon^{ab}\,F_{ab}^i\right) \right]\;,
\ee
with $\epsilon^{ab}=\epsilon^{abt}$. The operator $D$ is the covariant derivative for the restriction of $A$ to $\Sigma$. The interpretation is straightforward: the canonical pairs are formed by the components of the connection $A_a^i$ and their momenta $E^a_i = \delta_{ij}\epsilon^{ab}\,e_b^j$, the fundamental bracket being
\beq
\bigl\{A_a^i(x),E^b_j(y)\bigr\} = \delta_a^b\,\delta_j^i\,\delta^{(2)}(x-y)\;.
\ee
Furthermore, the time components $e_t$ and $A_t$ appear as Lagrange multipliers, enforcing the Hamiltonian to be a combination of constraints
\beq
D_aE^a_i = 0\;,\qquad
F_{ab}^i = 0\;. \label{flat constraint bf}
\ee
The first one is the Gau\ss{} law, which generates the $\SU(2)$ gauge transformations on the canonical variables. The second equation, $F_{ab}^i=0$, constrains the connection to be flat and generates the spatial part of the special symmetry \eqref{translation sym} via the brackets. Together, these constraints form an $\operatorname{ISU}(2)$ algebra.

When the frame field is non-degenerate, it is possible to extract from the multiplier $e_t$ the usual lapse and shift variables, so as to split the constraints $F_{ab}^i=0$ into a part generating diffeomorphisms of $\Sigma$ and another part called the scalar, or (somehow loosely) Hamiltonian constraint, \cite{thiemann-3d}. To this aim, we introduce the normal density vector
\beq \label{normal density}
n^i = \eps^{ijk}\,\eps_{ab}\,E^a_j\,E^b_k = \bigl(\vec{e}_1\times \vec{e}_2\bigr)^i\;,
\ee
using the vector notation for internal indices. Also define the curvature density vector, $F^i = \eps^{ab}F^i_{ab}$. It turns out that the norm of $\vec{n}$ is exactly the determinant of the 2-metric on $\Sigma$ induced by the co-triad: $\lvert \vec{n}\rvert^2 = \det({^2}g)$. Then, since $n^iE^a_i = 0$, we can write
\beq
\vec{e}_t\cdot\vec{F} = N^a\,V_a + \f{1}{2\sqrt{\det({^2}g)}}\,NH\;,
\ee
where $N^a$ and $N$ are respectively the shift and the lapse, defined as
\beq
N^a = \f{1}{\det({^2}g)}\eps^i_{\phantom{i}jk}\,E^a_i\,e_t^j\,n^k\;, \qquad N = \f{1}{\sqrt{\det({^2}g)}} \vec{e}_t\cdot\vec{n}\;.
\ee
This way the vector constraint $V_a$, which imposes the diffeomorphism invariance over $\Sigma$, and the scalar constraint $H$, which is simply the projection of $\vec{F}$ onto the normal $\vec{n}$, appear
\begin{align} \label{V}
V_a &= E^b_i\,F_{ab}^i = \vec{e}_a\cdot\vec{F}\;,\\
H &= \eps_{i}^{\phantom{i}jk}\,F^i_{ab}\,E^a_j\,E^b_k = \f{\vec{n}}{\lvert\vec{n}\rvert}\cdot\vec{F}\;. \label{H}
\end{align}

At this stage, an important remark is in order. In contrast with the proposal of \cite{thiemann-3d}, we will consider a Hamiltonian $H$ of density weight 2 rather than one. The reason for this choice is that we will perform the most naive regularization and quantization, in which $E^a_i$ and $F_{ab}^i$ are easily treated, but not non-analytic factors like $\f{1}{\sqrt{\det({^2}g)}}$.

\subsection{Loop quantization}

We now proceed to the loop quantization on the unreduced phase space, and we will then study the reduction by the constraints at the quantum level. First, the construction of the space of states in the connection polarization is based on cylindrical functions, which probe the connection only through a finite number of variables: the holonomies $U_e(A)$ of the connection $A$ along some set of edges $(e)$. Given a graph $\Gamma$, with $E$ edges and $V$ vertices, and a function $f$ over $(\SU(2))^E$, we form the cylindrical function $\psi_{\Gamma,f}$  as
\beq
\psi_{\Gamma,f}(A) \,=\, f\bigl( U_{e_1}(A),\dotsc, U_{e_E}(A)\bigr)\;.
\ee
We are moreover interested in $\SU(2)$ gauge invariant states. Gauge transformations act on holonomies only on their endpoints. If $h$ is a map from $\Sigma$ to $\SU(2)$, then: $U_e(A^h) = h(t(e))\,U_e(A)\,h(s(e))\mone$, with $t(e), s(e)$ being respectively the source and target points of the path $e$. When focusing on a single graph $\Gamma$, this reduces gauge transformations to an action of $\SU(2)^V$ on the set of cylindrical functions over $\Gamma$. So from any function $f$ over $\SU(2)^E$, one gets an invariant function by averaging over the $\SU(2)^V$ action.

The algebra of such functions has a natural inner product which comes from the Haar measure on $(\SU(2))^E$,  $d\mu_\Gamma = \prod_e dg_e$, giving
\beq \label{ps graphe}
\langle \phi_{\Gamma,h} | \psi_{\Gamma,f}\rangle \,=\, \int \prod_{e=1}^E dg_e\ \bar{h}(g_1,\dotsc,g_E)\,f(g_1,\dotsc,g_E)\;.
\ee
Completing this space for the corresponding norm leads to the Hilbert space $L^2\bigl(\SU(2)^E/\SU(2)^V, d\mu_{\Gamma}\bigr)$.

To get the full kinematical Hilbert space of loop quantum gravity, one has to consider a sort of union of cylindrical functions for all possible graphs. This results in a Hilbert space which can be understood as $L^2(\bar{\calA},d\mu_{\rm AL})$, the space of square-integrable functionals over $\bar{\calA}$. An element of $\bar{\calA}$ is a generalized connection, which assigns a holonomy to any path, while the Ashtekar-Lewandowski measure $\mu_{\rm AL}$, arising from the set of measures $\mu_\Gamma$, makes cylindrical functionals on different graphs orthogonal (see for instance \cite{ashtekar-status-report,thiemann-lectures-lqg}).


The spin network basis is an orthogonal basis of $L^2\bigl(\SU(2)^E/\SU(2)^V, d\mu_{\Gamma}\bigr)$. The first idea is to expand any element of $L^2(\SU(2)^E,d\mu_\Gamma)$ onto the matrix elements of the irreducible representations of $\SU(2)^E$, labelled by $E$ spins $j_e\in\N/2$. Then, rotation invariance at vertices imposes to contract the matrix elements of edges meeting at a vertex with intertwiners $\iota_v$. If $v$ is a node of $\Gamma$ with ingoing links $(e_{\rm in})$, and outgoing links $(e_{\rm out})$, then an intertwiner $\iota_v$ is a map: $\otimes_{e_{\rm in}}\calH_{j_e} \rightarrow \otimes_{e_{\rm out}}\calH_{j_e}$, (where $\calH_j$ is the carrier space of the spin $j$ representation) which commutes with the group action. Thus, to form the spin network basis on $\calH_\Gamma$, we need to assign spins to edges and a basis of intertwiners at each node. For a trivalent vertex with ingoing edges for example, there is a single intertwiner $\iota_{j_1 j_2 j_3}: \calH_{j_1}\otimes \calH_{j_2} \otimes \calH_{j_3} \mapsto \C$ up to normalization, its components in the standard magnetic number basis being the Wigner 3jm-symbol $\langle j_1, m_1 ; j_2, m_2 ; j_3, m_3 | \iota_{j_1 j_2 j_3}| 0\rangle = \left(\begin{smallmatrix} j_1 &j_2 &j_3 \\m_1 &m_2 &m_3 \end{smallmatrix}\right)$. Hence, spin network states are the following $\SU(2)^V$-invariant functions
\beq \label{def spinnet}
s_\Gamma^{\{j_e,\iota_v\}}(g_1,\dotsc,g_E) \,=\, \sum_{\{m_e, n_e\}} \prod_{e=1}^E \langle j_e, m_e\lvert g_e\rvert j_e, n_e\rangle\ \prod_{v=1}^V \langle \otimes_{e\,{\rm in}}\, j_e,m_e \lvert \iota_v \rvert \otimes_{e\,{\rm out}}\, j_e, n_e\rangle\;,
\ee
which form an orthogonal set. This enables a transform between gauge invariant functions on $\Gamma$ and functions over the colorings $(j_e,\iota_v)$.

Let us now look at the algebra of operators. Quite clearly, spin network functions form an algebra, so that a gauge-invariant holonomy operator can be defined as acting on $\calH_\Gamma$ by simple multiplication. In order to form quantities depending on the triad $E^a_i$, it is natural to smear it along curves, thus defining the fluxes
\beq
X_{c,f} = \int_c E^a_i\,f^i\ \eps_{ab}\,dx^b\;,
\ee
for a path $c$ and a function $f$. This is the natural choice given the chosen smearing of the connection so that the Poisson bracket closes on the space of cylindrical functions (see typically \cite{corichi-zapata-fluxes}). It can be shown that the bracket $\{ X_{c,f},\psi_{\Gamma,h}(A)\}$ receives contributions only from the crossings of $c$ with $\Gamma$, and that there the flux act as a left or right invariant vector field on the function $h$.

This framework allows to define different kinematical observables such as the lengths of curves, and areas, like in \cite{livine-3dlength}. As we will need some of these, we will proceed to a simplified derivation based on the phase space of loop quantum gravity restricted to a single graph, section \ref{sec:quantum}.

\subsection{The flat state on the boundary of a tetrahedron}

For concreteness of the discussion, we consider the 2-sphere as our canonical surface, triangulated with four triangles which are glued like on the boundary of a tetrahedron. Our computation actually works without amendments for any 3-valent vertex of a triangulation. It can also be extended to vertices of higher valence (by duality, to cycles of spin network graphs containing an arbitrary number of links)\footnote{Though, the recurrence relations one gets in these generalized situations are more intricate, and require further discussion beyond the scope of the present note (see \cite{recurrence-paper}).}.

\begin{figure}\begin{center}
\includegraphics[width=4.5cm]{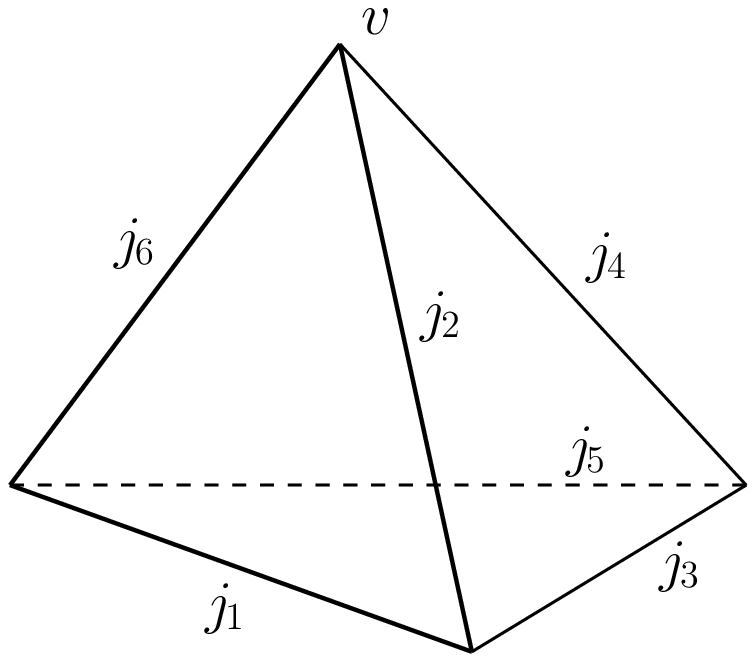}
\caption{ \label{fig:tet}
}
\end{center}
\end{figure}

The kinematical Hilbert space of gauge invariant states is spanned by spin network states on the oriented graph dual to the triangulation, which is also a tetrahedral graph. The states can be seen as $L^2$ functions over the six $\SU(2)$ elements on the dual edges and invariant under translation acting at the four dual vertices. Let us write explicitly the spin network function, labelled by six spins $(j_1,\dotsc,j_6)$, following the figure \ref{fig:tet},
\begin{multline} \label{defspinnet}
s^{\{j_e\}}_{\rm tet}(g_1,\dotsc,g_6) = \sum_{\substack{ a_1,\dotsc,a_6 \\b_1,\dotsc,b_6}} \begin{pmatrix} j_1 &j_2 &j_3\\ b_1 &-a_2 &-a_3\end{pmatrix} \begin{pmatrix} j_1 &j_5 &j_6\\ -a_1 &b_5 &-a_6\end{pmatrix} \begin{pmatrix} j_3 &j_4 &j_5\\b_3 &-a_4 &-a_5\end{pmatrix} \begin{pmatrix} j_2 &j_6 &j_4\\ b_2 &b_6 &b_4\end{pmatrix} \\
\times \left[\prod_{e=1}^6 (-1)^{j_e-a_e} \langle j_e, a_e \vert g_e \vert j_e,b_e\rangle\right] \;.
\end{multline}
The range of summation is $-j_e\leq a_e,b_e\leq j_e$ for each link. In the sum, each link $e$ carries two magnetic indices, $b_e$ on the source vertex which is contracted with the right of the holonomy $g_e$, and $a_e$ on the target vertex and contracted with the left of $g_e$.

Given the trivial topology we consider, there is only one physical state, which is not normalizable. It formally corresponds to the product of Dirac deltas for the holonomies around the independent cycles of the graph. Cycles can be defined as the boundaries of the faces dual to the vertices of the triangulation. The product has to be considered over independent cycles, which in the present case is easy to understand: asking for the holonomies around three faces to be trivial enforces the same condition on the remaining, fourth face. So we choose any three cycles like (see picture \ref{fig:tet})
\beq
\psi_{\operatorname{phys}}(g_1,\dotsc,g_6) = \delta(g_4\mone g_5 g_6)\,\delta(g_2\mone g_1\mone g_6)\,\delta(g_2\mone g_3 g_4)\;.
\ee
To get its expansion on the spin network basis, we take the inner product with a spin network function $s^{\{j_e\}}_{\rm tet}(g_1,\dotsc,g_6)$,
\beq
\psi_{\operatorname{phys}}(j_1,\dotsc,j_6) = \int \prod_{e=1}^6 dg_e\ s^{\{j_e\}}_{\rm tet}(g_1,\dotsc,g_6)\ \psi_{\operatorname{phys}}(g_1,\dotsc,g_6) = \begin{Bmatrix} j_1 &j_2 &j_3 \\j_4 &j_5 &j_6 \end{Bmatrix}\;.
\ee
The quantity into brackets is known as the Wigner 6j-symbol. This result simply comes from the fact that the physical state enforces the evaluation of the integral on the flat connections (up to gauge transformations), while there: $s_{\{j_e\}}(\mathbbm{1})=\{6j_e\}$.

The 6j-symbol is known to satisfy the Biedenharn-Elliott, or pentagon, identity
\begin{equation} \label{bied-elliott}
\begin{Bmatrix} j &h &g\\ k &a &b \end{Bmatrix}\,\begin{Bmatrix} j &h &g\\ f &d &c \end{Bmatrix} = \sum_l (-1)^{S+l}(2l+1)\,\begin{Bmatrix} k &f &l\\ d &a &g\end{Bmatrix}\,\begin{Bmatrix} a &d &l\\ c &b &j\end{Bmatrix}\,\begin{Bmatrix} b &c &l\\ f &k &h\end{Bmatrix}\;,
\end{equation}
where $S$ is the sum of the nine spins. It is key to the Ponzano-Regge state-sum model, since it corresponds to stating its invariance under a change of the spacetime triangulation by a 2-3 Pachner move\footnote{To prove topological invariance, one also needs to check the invariance under the 1-4 Pachner move. It can be done using again the Biedenharn-Elliott identity and some orthogonality properties of the 6j-symbol. However, the latter invariance is only formal since it results in a divergent sum, \cite{ooguri-3d, twisted-bubbles}.}.

Specializing \eqref{bied-elliott} to the case $f=1$, one is free to choose $g=d$ and $h=c$, while the sum on the right hand side reduces to only three terms, for $l=k-1,k,k+1$. This leads to the second order recurrence relation
\be
A_{+1}(j_1)\,\begin{Bmatrix} j_1+1 &j_2 &j_3 \\ j_4 &j_5 &j_6 \end{Bmatrix} +
A_{0}(j_1)\,\begin{Bmatrix} j_1 &j_2 &j_3 \\ j_4 &j_5 &j_6 \end{Bmatrix} +
A_{-1}(j_1)\,\begin{Bmatrix} j_1-1 &j_2 &j_3 \\ j_4 &j_5 &j_6 \end{Bmatrix} = 0\;.
\ee
The coefficients are
\begin{align} \label{A_0}
A_0(j_1) &= (-1)^{j_2+j_4+j_6}\begin{Bmatrix} j_2 & j_2 &1 \\ j_6 &j_6 &j_4\end{Bmatrix} + (-1)^{2j_1+j_2+ j_3+j_5+j_6}(2j_1+1)\ \begin{Bmatrix} j_1 & j_1 &1\\ j_2 &j_2 &j_3\end{Bmatrix}\,\begin{Bmatrix} j_1 & j_1 &1\\ j_6 &j_6 &j_5\end{Bmatrix}\;,\\
A_{\pm 1}(j_1) &= (-1)^{2j_1+j_2+ j_3+j_5+j_6+1}\bigl(2(j_1\pm1)+1\bigr)\ \begin{Bmatrix} j_1\pm1 & j_1 &1\\ j_2 &j_2 &j_3\end{Bmatrix}\,\begin{Bmatrix} j_1\pm 1 & j_1 &1\\ j_6 &j_6 &j_5\end{Bmatrix}\;. \label{Apm1}
\end{align}
They can obviously be evaluated explicitly (but the above expressions will be more useful to us), and the recurrence relation can be solved, starting from an initial condition to determine all values of the 6j-symbol.

An interesting feature is that though it is a second order difference equation (involving three consecutive neighbors), the solution is fully determined by a single initial condition. Looking at the lowest possible value of the spin $j_1$, that is: $j_1^{\min} = \max(\vert j_2-j_3\vert, \vert j_5-j_6\vert)$, it turns out that the lowering coefficient $A_{-1}(j_1^{\min})$ evaluated on this spin is zero. Thus, the recurrence can be implemented starting from the initial value on this lowest spin\footnote{This property is important with respect to the asymptotic limit. There, it is known \cite{varshalovich-book, schulten-gordon2, maite-etera-6jcorr} that the 6j-symbol oscillates like the cosine of the Regge action $S_{\rm R}(j_e) = \sum_e (j_e+\f12)\theta_e$, for a tetrahedron with lengths $(j_e+\f12)$,
\beq
\begin{Bmatrix} j_1 &j_2 &j_3\\j_4 &j_5 &j_6\end{Bmatrix} \approx \f1{\sqrt{12\pi V(j_e)}}\,\f12\,\Bigl(e^{i(S_{\rm R}(j_e)+\f\pi4)} + e^{-i(S_{\rm R}(j_e)+\f\pi4)}\Bigr)\;.
\ee
Having a second order equation, it may have been expected to be solved in the asymptotics by both the positive and negative exponentials of the action. But the model is fully independent from the orientation: there is only one required initial condition, because of the vanishing of $A_{-1}(j_1^{\min})$, leading in the large spin regime to the cosine of the action.}.

So, the recurrence relation and the Biedenharn-Elliott identity
\begin{itemize}
 \item encode the invariance properties of the Ponzano-Regge model, or equivalently they are responsible for its special symmetries,
 \item and fully determine the 6j-symbol which is the physical state in the spin network basis for the triangulated 2-sphere.
\end{itemize}
To make these two aspects talk to each other, we would like to derive the recurrence relation as an equation encoding the symmetries in the context of loop quantum gravity, in the form of \eqref{wdw eq}.

\section{The proposal for a new Hamiltonian} \label{sec:class}

\subsection{The discrete geometries of loop quantum gravity on a single graph}

When looking at a graph, the loop quantization, which is based on specific smearings of the fields, can be reproduced starting from a classical phase space associated to the graph. The geometric interpretation of this phase space will be of particular importance for us.

Let us restrict attention to a single oriented graph, typically the dual to a triangulation of $\Sigma$. The phase space inherited from the smearings which are used in loop quantum gravity is the cotangent bundle $T^*(\SU(2))^E$. More precisely, the pairs of canonical variables $(A_a^i,E^a_i)$ are replaced with a pair $(g_e, X_e)\in\SU(2)\times \su(2)$ on each edge of $\Gamma$. The group element $g_e$ is the parallel transport operator along $e$ from its source vertex to its target vertex, while $X_e$ is the flux variable for an edge dual to $e$. We chose $X_e$ to act as a right derivative, which means that we have the brackets
\beq
\bigl\{X_e^i,X_e^j\bigr\} = \eps^{ij}_{\phantom{ij}k}\,X_e^k\;,\qquad \bigl\{X_e^i, g_e\bigr\} = g_e\,\tau^i\;,
\ee
where the matrices $(\tau_i)_{i=1,2,3}$ are anti-hermitian matrices\footnote{They read: $\tau^i = -\f{i}{2}\sigma^i$, in terms of the Pauli matrices $(\sigma^i)$.} satisfying $[\tau^i,\tau^j] = \eps^{ij}_{\phantom{ij}k}\tau^k$. All other brackets vanish. The right derivatives are obtained by transporting $X_e$ to the target vertex of $e$,
\beq \label{Xtilde}
\widetilde{X}_e = \Ad(g_e)\, X_e\;, \quad \text{together with} \quad
\bigl\{\widetilde{X}_e^i,\widetilde{X}_e^j\bigr\} = -\eps^{ij}_{\phantom{ij}k}\,\widetilde{X}_e^k,\qquad \bigl\{\widetilde{X}_e^i, g_e\bigr\} = \tau^i\,g_e\;.
\ee
Here $\Ad$ is the standard adjoint action of the group on its algebra, and in particular, it is here an action of holonomies on fluxes.

In this framework, the Gau\ss{} law becomes the sum of the fluxes meeting at a vertex $v$ of $\Gamma$,
\beq \label{closure}
\sum_{e \supset v} X_e = 0\;,
\ee
if all edges are outgoing.Using this equation, we can get a nice interpretation of the flux variables as describing the discrete intrinsic geometry of the graph dual to $\Gamma$. For instance if $\Gamma$ is dual to a triangulation of $\Sigma$, it is natural to see the variables $X_e$ as 3-vectors describing the edges of the triangulation in $\R^3$, so that the above equation is nothing but the closure of each triangle.

One can then extract the intrinsic geometry as follows. The length of an edge $e^*$ of the triangulation is the norm of $X_e$, for $e$ being the dual link to $e^*$, $\ell_{e} = \vert X_e\vert$. The dihedral angle $\phi_{ee'}$ between two edges $(e^*, e^{'*})$ of a triangle is contained in the dot product of $X_{e}, X_{e'}$, where the dual edges $e, e'$ meet at a vertex,
\beq
\cos \phi_{ee'} = -\eps_{ee'}\ \f{X_e\cdot X_{e'}}{\vert X_e\vert \vert X_{e'}\vert}\;,
\ee
where $\eps_{ee'}$ is $1$ if $e$ and $e'$ are both outgoing or ingoing, and $-1$ else.

It is also possible to introduce some notions of 3d dihedral angles, i.e. between two triangles sharing an edge of the triangulation. Since it will be a central point in our discussion on the dynamics, we postpone its presentation to the next section.

\subsection{Construction of the new Hamiltonian}

The usual flatness constraint on a triangulation, inspired from Regge calculus, is to consider curvature concentrated around $(d-2)$-simplices, i.e. here vertices of the triangulation. In the dual picture this corresponds to looking at the parallel transport operators around the faces of $\Gamma$, and ask them to be trivial, so that
\beq \label{3d canonical simplicial flatness}
g_f = \prod_{e\subset f} g_e = \mathbbm{1}\;.
\ee
Together with the Gau\ss{} law \eqref{closure}, they form a system of first class constraints, in which the above flatness constraint generates translations in $\R^3$ of the vertices of the triangulation. This is the usual starting point when one wants to reduce the phase space before quantization, since it is easy to see that the theory has actually only a finite number of degrees of freedom (given in terms of the topology of $\Sigma$).

This is also the form of the constraint which has been mainly used to reduce after loop quantization and make the link with spin foams, like in \cite{ooguri-3d,noui-perez-ps3d}. From the continuum LQG point of view, the underlying idea is to define the projector onto flat connections through its matrix elements between any two spin network states. This is done using a lattice regularization which is adapted to the spin network graphs one wants to consider, like a 3-valent graph $\Gamma$, dual to a triangulation and which contains the spin network graphs. Then, one has to define the flatness constraint on plaquettes of $\Gamma$, \eqref{3d canonical simplicial flatness} for instance, and this results in the projection of any spin network whose graph is contained in $\Gamma$ onto its restriction to flat connections. When evaluating physical inner products, the topological nature of the moduli space of flat connections guarantees the independence with respect to the triangulation used to regularize the constraints. Thus, it can be checked that the regulator can be removed \cite{noui-perez-ps3d}. We will use implicitly this reasoning and hence simply work on a fixed graph $\Gamma$, dual to a triangulation. Since we decide not to use \eqref{3d canonical simplicial flatness}, we have to see if our approach makes us able to recover the projector onto physical states on $\Gamma$. This will be done partially in the section \ref{sec:quantum} for trivalent faces.

In contrast to the form \eqref{3d canonical simplicial flatness}, we are here mostly interested in a form of the constraints which could be closer to what we need for 4d gravity, like in \cite{thiemann-3d}, and at the same time make contact with recurrence relations which are a way to encode the symmetries at the quantum level in spin foams \cite{recurrence-paper}. Our proposal is to quantize the following operator in the spin network basis
\beq \label{lqg hamiltonian}
H_{v,f} = X_{e_1}\cdot X_{e_2} - X_{e_1}\cdot \Ad(g_f)\bigl(X_{e_2}\bigr) = \sum_{i,j=x,y,z} X_{e_1}^i\,\bigl( \delta_{ij} - R(g_f)_{ij}\bigr)\,X_{e_2}^j\;.
\ee
and show that the corresponding constraint equation is exactly the recursion relation on the 6j-symbol when the triangulation is the boundary of a tetrahedron. Here $e_1, e_2$ meet at the vertex $v$ of $\Gamma$, and are in the boundary of the face $f$. In the first line, we have used the adjoint action of the group on its algebra, and in the second line, we have emphasized that it is equivalent to the vector (spin 1) representation, given in the Cartesian basis by the usual rotation matrices.

First we give a nice geometric interpretation of this quantity in terms of discrete geometries, which in our view strengthens and makes more precise some of the ideas of \cite{dittrich-ryan-simplicial-phase, dittrich-simplicity} which were proposed at the classical level to describe a flat 4-simplex from its boundary.

There are two basic ideas which support our choice for $H_{v,f}$. The first key idea is to represent the scalar constraint $H$, in the following way. As can be seen in \eqref{H}, the constraint is just the contraction of the curvature with the normal $\vec{n}$, \eqref{normal density}, to the canonical surface. Given the choice of smearings in LQG, the curvature on a spin network graph is naturally regularized as the product of the holonomies along the cycles of the graph, where a cycle is defined as the boundary of a face delimited by $\Gamma$ on the canonical surface. In practice the curvature tensor is replaced with
\beq
\epsilon^{ij}_{\phantom{ij}k}\,F_{ab}^k \quad \longrightarrow\quad \delta^{ij} - \bigl(R(g_f)\bigr)^{ij}\;.
\ee
Equivalently, the curvature is the effect of parallel transport around a vertex of the dual graph to $\Gamma$. But then, it is quite unclear what a normal to this vertex means. So we choose a face touching this vertex and consider the normal to this face. Back to the description on the spin network graph, the latter is dual to a vertex of $\Gamma$, and choosing a normal $N_t$ is equivalent to choosing two flux variables $X_{e_1}$ and $ X_{e_2}$ meeting there and both lying in the boundary of a common face, with: $N_t = X_{e_1}\times X_{e_2}$. So the natural choice is to contract the regularized curvature along a cycle of $\Gamma$ with $X_{e_1}$ and $ X_{e_2}$, where $e_1, e_2$ are edges of the given cycle $c=\partial f$ meeting at a vertex $v$,
\beq
E^a_i\,E^b_j \quad \longrightarrow\quad X_{e_1}^i\,X_{e_2}^j\;.
\ee
Also, we choose $v$ as the base point to define $g_f$ and take its orientation so that it transports first along $e_2$ and ends with $e_1$. Then
\beq
E^a_i\,E^b_j\ \epsilon^{ij}_{\phantom{ij}k}\,F_{ab}^k\quad \longrightarrow\quad H_{v,f}\;.
\ee

One may ask then about the splitting of the constraint \eqref{3d canonical simplicial flatness} into scalar and diffeomorphism constraints and the precise status of our proposal in this respect. Remember this splitting comes from projecting the curvature onto the normal $\vec{n}$ to $\Sigma$ on the one hand, \eqref{H}, and onto the directions tangent to $\Sigma$ on the other hand, \eqref{V}. Here assuming we are dealing with a triangular face $f$, we have three different $H_{v,f}$, one for each vertex. Each of them corresponds to choosing the normal to one of the three triangles meeting at the node dual to $f$. So for generic $X_e$, the whole set of $H_{v,f}$ for a given face takes into account the projection of the holonomy around $f$ in the three independent directions of $\R^3$. For this reason, there is no need to introduce diffeomorphism constraints in this framework.

A crucial fact to make our assertion complete is to check that the algebra of the set of $(H_{v,f})$ on $\Gamma$ closes. Here comes the second key idea of our construction. If one wants to impose the flatness constraint as a Wheeler-DeWitt equation, one needs to find three (real) independent constraints on the matrix elements of the holonomy $g_f$. To do it in a gauge invariant way, one has to project the matrix $R(g_f)$ onto the fluxes, and this is how $H_{v,f}$ is constructed.

Consider a cycle with three edges, say $e_1, e_2, e_6$ like in the figure \ref{fig:tet}, with holonomy $g_f = g_6\mone g_1 g_2$. The three functions $H_{v,f}$ on this cycle are actually the three constraints obtained by contracting $(\id - R(g_f))$ with pairs formed among three vectors defined at the node $v$ where $e_2, e_6$ meet, which $X_2,X_6$ and $X_1(v)$, where $X_1(v) = \Ad(g_2\mone)X_1$. We denote the Hamiltonians by $H_{26}, H_{21}, H_{16}$. For example, $H_{21} = X_2\cdot (\id - R(g_f))X_1(v)$.

In this sense, this is just looking at the matrix elements of $R(g_f)$ in a non-orthogonal basis. For a generic triangulation, there is one constraint for each node of a cycle. Thus there are the appropriate number of constraints to enforce $R(g_f)=\id\in\SO(3)$, when the flux variables around the cycle span the three dimensions of the algebra.

To complete the argument, we first show that the three $H_{v,f}$ around $f$ are good coordinates in the neighborhood of the solution\footnote{Notice that the solutions of the equation $X_2\cdot X_6 - X_2\cdot \Ad(g) X_6=0$ are $g= \exp(t_2 X_2)\exp(\eta (X_2\times X_6)) \exp(t_6 X_6)$, where $t_1, t_2$ are arbitrary. But $\eta$ admits only a finite number of values, since there is a finite number of $\SO(3)$ rotations with axis $X_2\times X_6$ which solve the constraint. It seems that non-zero $\eta$ however only survives if $X_2\times X_6$ is proportional to the third vector $\Ad(g_2\mone)X_1$.}$R(g_f)= \id$. Writing $g(\vec p) = \sqrt{1-p^2}\,\unit + i\,\vec p\cdot \vec \sigma \in \SU(2)$, we have
\beq
X\cdot X' -X\cdot R(g)X' = -2\sqrt{1-p^2}\,\bigl(X\times X'\bigr)\cdot \vec p + 2\bigl[ p^2\bigl(X\cdot X'\bigr) - \bigl(X\cdot\vec p\bigr)\bigl(X'\cdot \vec p\bigr)\bigr]\;.
\ee
From this one finds $\vert dH_{26}\wedge dH_{21}\wedge dH_{16}\vert_{R(g_f)=\id} = 8\vert \det ( X_2, X_6, X_1(v))\vert^2$, where the differential is taken with respect to $\vec p_f$. Hence the system is generically of rank three, and this makes clear why we need the three vectors to span a basis of $\R^3$.

Finally, all Poisson brackets weakly vanish. Consider the bracket between $H_{12}$ and $H_{16}$ which are in the same face. It reads
\beq
\left\{ H_{12}, H_{16}\right\} = \bigl(X_6\times X_2\bigr)\cdot \bigl[\id - R(g_f)\bigr] X_1(v) + \bigl(X_2+X_6\bigr)\cdot \bigl(X_1(v)\times R(g_f)X_1(v)\bigr)\;,
\ee
which is zero when $R(g_f)= \id$. Consequently it can be re-expressed as a linear combination of $H_{26}, H_{21}, H_{16}$.

On the boundary of the tetrahedron (figure \ref{fig:tet}), the other non-trivial brackets are $\{H_{12}, H_{23}\}$, $\{H_{12}, H_{34}\}$, $\{H_{12}, H_{24}\}$ and $\{H_{12}, H_{64}\}$ (and the other are deduced by symmetry, or identically zero). It is straightforward to check that they can be written with the quantities $(\id - R(g_f))$ and $X_e\times R(g_f)X_e$.

\subsection{Its geometric meaning}

\begin{figure}\begin{center}
\includegraphics[width=5cm]{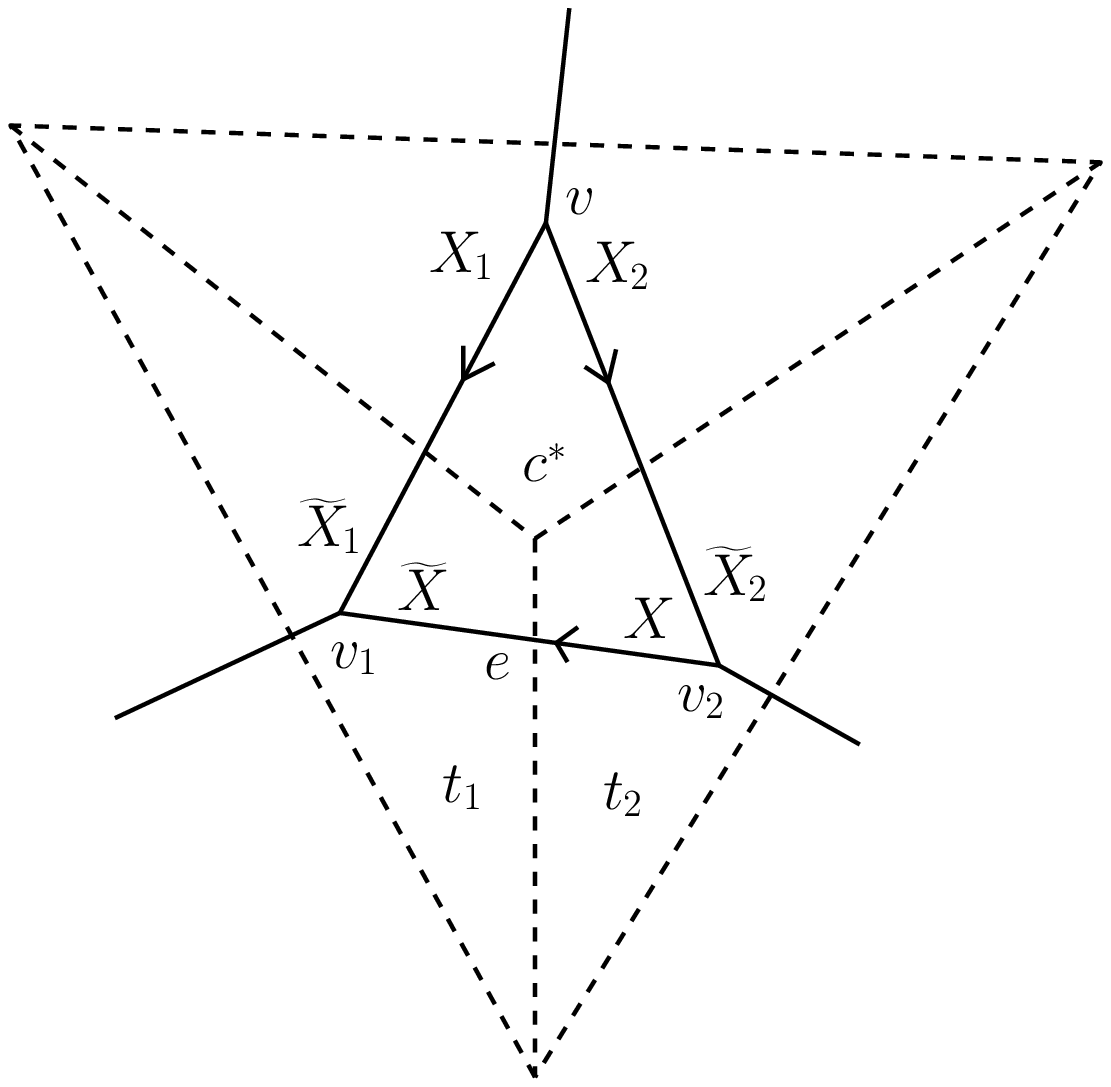}
\caption{ \label{fig:3-cycle}
}
\end{center}
\end{figure}

Physically, what we expect from the classical flatness constraint $F_{ab}^i=0$ is that $\Sigma$ with its geometry can be locally embedded into flat 3-space. From the point of view of the geometry of a triangulation, it means that the (3d) dihedral angles $\Theta_{t_1 t_2}$ between two adjacent triangles $t_1, t_2$ should be given as a standard function of the dihedral 2d angles $(\phi_{ee'})$. A typical situation is depicted at figure \ref{fig:3-cycle}, for a trivalent node of the triangulation. The triangles $t_1, t_2$ are dual to the vertices $v_1, v_2$ of the spin network graph $\Gamma$, which are linked by the edge $e$. The classical angle between the two triangles $t_1, t_2$ can be computed from the three 2d angles around the node $c^*$, which is dual to the cycle $c$, using the formula
\beq \label{regge flat}
\cos \Theta_{t_1 t_2}\,(X,\tl{X}) = \f{\cos\phi_{e_1 e_2} - \cos\phi_{e_1 e}\,\cos\phi_{e_2 e}}{\sin\phi_{e_1 e}\ \sin\phi_{e_2 e}}\;.
\ee
We have written it as a function of the flux variables $X_e, \tl{X}_e$ since all angles $\phi_{ee'}$ can be evaluated from them, without writing down holonomies. In addition, when the closure relation \eqref{closure} holds classically, the angles $\phi_{ee'}$ are determined as functions of the lengths $(\ell_e)$, and hence so are the angles $\Theta_{t_1t_2}$.

Parallelly, using the interpretation of the flux variables and the holonomies, there is a natural notion of dihedral angles between triangles, defined as follows. Since the fluxes represent the directions of the edges of the triangulation, the normal $N_t$ to a triangle is given by the wedge product of two of them. Following picture \ref{fig:3-cycle}, $N_{t_1}^i = \epsilon^i_{\phantom{i}jk}\, \tl{X}_e^j\, \tl{X}_{e_1}^k$, and $N_{t_2}^i = \epsilon^i_{\phantom{i}jk}\,\tl{X}_{e_2}^j\,X_e^k$. It suggests to look at the dot product of these normals as being the cosine of the dihedral angle between $t_1$ and $t_2$. But for this, it is necessary to transport them in a common frame, say $N_{t_2}$ along the edge $e$ to the vertex $v_1$. So we define the angle $\theta_{t_1t_2}$ as
\beq
\cos \theta_{t_1 t_2}\,(X, \tl{X}, g) = -\frac{N_{t_1}\cdot \Ad(g_e) N_{t_2}}{\vert N_{t_1}\vert\ \vert N_{t_2}\vert}\;.
\ee
This is a function of the holonomy $g_e$ since one has to compare the local embedding of $t_1$ to that of $t_2$. The result can be partially evaluated in terms of the 2d dihedral angles, because it is a dot product of two vector products in $\R^3$
\beq
\cos \theta_{t_1 t_2}\,(X, \tl{X}, g) = \f{\tl{X}_1\cdot \Ad(g_e)\tl{X}_2 \ -\ \cos\phi_{e_1 e}\,\cos\phi_{e_2 e}}{\sin\phi_{e_1 e}\ \sin\phi_{e_2 e}}\;.
\ee
Quite clearly, the dependance on the holonomies cannot be dropped off. Obviously, one would like to compare this notion of dihedral angle to the standard formula \eqref{regge flat} for the angle $\Theta_{t_1 t_2}$. It follows from the definition $\tl{X}_{e_1} = \Ad(g_{e_1})X_1$ (and the same for $e_2$) that the difference between $\theta_{t_1 t_2}$ and $\Theta_{t_1 t_2}$ is precisely $H_{v,f}$,
\beq
H_{v,f} = \sin\phi_{e_1 e}\ \sin\phi_{e_2 e}\ \Bigl( \cos \Theta_{t_1 t_2} \,-\, \cos \theta_{t_1 t_2}\Bigr)\;.
\ee

We will not go further about the classical aspects, though a full classical analysis would certainly be worth. It could be done either with the variables $(g,X)$ or with the gauge invariant variables $(l, \theta)$. In this respect, it is interesting to note (see \cite{waelbroeck-zapata, kadar}) that the angle $\theta_{e}=\theta_{t_1 t_2}\,(X, \tl{X}, g) $ associated to the edge $e$ is in fact the canonical variable dual to the edge length,
\be
\{l_{e},\theta_{e'} \} = \delta_{e,e'}\;.
\ee

\section{A Wheeler-DeWitt equation for the flat state on the boundary of a tetrahedron} \label{sec:quantum}

Now we look at the quantization of $H_{v,f}$ on the boundary of a tetrahedron, see figure \ref{fig:tet}, for $v$ the vertex where $j_2, j_4, j_6$ meet and $c=\partial f$ the cycle $(j_1 j_2 j_6)$,
\beq
H_{v=(e_2\cap e_6),f = (j_1 j_2 j_6)} \,=\, X_2\cdot X_6 \,-\, X_2\cdot \Ad(g_2\mone g_1\mone g_6)(X_6)\;.
\ee
This is done is two parts: first the quantization of the operator $(X_2\cdot X_6)$, which is basically the cosine of the angle between the edges $e_2, e_6$, and then the same quantity computed after parallel transport of $X_6$ around the face. In both cases, we use the space of gauge-invariant states $\calH_\Gamma$, spanned by spin network functions.

\subsection{Quantization of the cosine of the 2d dihedral angle}

The observables of the intrinsic geometry, lengths and 2d angles, are easily defined on spin networks. First\footnote{The scalar product is taken to be $X\cdot Y\equiv -X^{i}Y_{i}$ because we are using anti-hermitian generators.},
\beq
\bigl(\widehat{X_{e}^2}\,s_\Gamma^{\{j_e,\iota_v\}}\bigr)(g_1,\dotsc,g_E) = - \Bigl\{ X_e^i,\bigl\{ X_{e i},s_\Gamma^{\{j_e,\iota_v\}}(g_1,\dotsc,g_E)\bigr\}\Bigr\} = j_e\bigl(j_e+1\bigr)\ s_\Gamma^{\{j_e,\iota_v\}}(g_1,\dotsc,g_E)\;, \label{length op}
\ee
for the lengths, diagonalized in this basis with a discrete spectrum.

As for the dot product of different $X_e$ meeting at a node, it can evaluated as follows. We consider the edges $e_2, e_6$, meeting at a 3-valent node with $e_4$, and assume without loss of generality that they are all outgoing. Then, the product $X_2^i X_{6 i}$ acts by inserting $\tau_{e_2}^i\otimes\tau_{e_6 i}$ in the spin network, thus modifying the intertwiner. This corresponds to adding a link in the spin 1 representation between the two edges, on the right of the group elements $g_{e_2}, g_{e_6}$. In equations, we have
\beq \label{insertion x1dotx2}
\begin{split}
&\widehat{\bigl(X_2\cdot X_6\bigr)}\ \sum_{a_2, a_4, a_6} \begin{pmatrix} j_2 &j_6 &j_4 \\ a_2 &a_6 &a_4\end{pmatrix}\ g_{e_2}\rvert j_2, a_2\rangle \otimes g_{e_6}\rvert j_6, a_6\rangle \otimes g_{e_4}\rvert j_4, a_4\rangle\\
&= -\sum_{a_2, a_4, a_6} \begin{pmatrix} j_2 &j_6 &j_4 \\ a_2 &a_6 &a_4\end{pmatrix}\ g_{e_2}\,\tau^i\rvert j_2, a_2\rangle \otimes\ g_{e_6}\,\tau_i\rvert j_6, a_6\rangle \otimes\ g_{e_4}\rvert j_4, a_4\rangle\;.
\end{split}
\ee
To extract from this formula some Wigner recoupling coefficients (a 6j-symbol actually), we need a precise calculation of the so-called grasping $\tau_{e_2}^i\otimes\tau_{e_6 i}$.

It is convenient to go to a different basis of the Lie algebra, in which the flux variables $X_e^m$ have magnetic indices $m=-1,0,1$ instead of the Cartesian indices ($i=x,y,z$), and with generators $L_m$. A way to choose well adapted generators is the following. We have used so far the adjoint action of holonomies on fluxes, which is equivalent to the spin 1 representation. But at the quantum level, the matrix elements of the adjoint action act by multiplication on cylindrical functions. Because we built the spin network functions in terms of Wigner matrices, it is natural to choose a basis where the matrix elements of the adjoint representation are the Wigner matrices with spin 1.

We can test this request on the parallel transport relation \eqref{Xtilde} at the quantum level. In the Cartesian basis, we have: $(\Ad(g_e) X_e)^i = \sum_j R(g_e)^i_{\phantom{i}j} X_e^j$, where $R(g)$ is the standard rotation matrix. When acting on matrix elements (of the spin network basis), that becomes
\beq
\sum_j R(\widehat{g}_e)^i_{\phantom{i}j}\, \widehat{X}_e^j\ \langle j,b\vert\, g_e\,\vert j, a\rangle = \sum_j R(g_e)^i_{\phantom{i}j}\ \langle j,b\vert\, g_e\ \tau^j\,\vert j, a\rangle = \langle j,b\vert\, \tau^i\ g_e\,\vert j, a\rangle = \widehat{\widetilde{X}}_e^i\ \langle j,b\vert\, g_e\,\vert j, a\rangle\;.
\ee
We have used the well-known intertwining property: $R(g)^i_{\phantom{i}j}\tau^j = g\mone \tau^i g$. Thus, the classical relation is correctly implemented on wave functions. We now ask for the same relation to hold in the new basis $(L_m)$ and with the Wigner matrix of spin 1 instead of the rotation matrix $R(g)$,
\beq
\langle j,b\vert\, g_e\mone\,L_m\,g_e\,\vert j,a\rangle = \sum_{n=-1,0,1} \langle 1,m\vert\,g\,\vert 1,n\rangle \ \langle j,b\vert\,L_n\,\vert j,a\rangle\;.
\ee
That shows that the matrix-valued vector $L$ in the spin $j$ is actually an intertwiner $\calH_{j}\otimes \calH_{j^*}\otimes \calH_1\rightarrow \C$, where $\calH_{j^*}$ is the dual representation. Hence, the element $\langle j,b\vert\,L_m\,\vert j,a\rangle$ must be proportional to the Wigner 3mj-symbol $(-1)^{j-a}\left(\begin{smallmatrix} 1 &j &j\\ m &-a &b\end{smallmatrix}\right)$. We choose the normalization\footnote{Notice that these generators are quite unusual since
\beq
L_0\vert j,a\rangle = a\,\vert j,a\rangle, \quad \text{and}\qquad L_\pm\,\vert j, a\rangle = \mp \sqrt{\f{j(j+1)-a(a\mp1)}{2}}\,\vert j, a\mp1\rangle\;.
\ee} like
\beq
\langle j,b\vert\,L_m\,\vert j,a\rangle = (-1)^{2j+1}\,N_j\ (-1)^{j-a}\begin{pmatrix} 1 &j &j\\ m &-a &b\end{pmatrix}\;,
\ee
the prefactor being
\beq
N_j = \sqrt{j(j+1)\,d_j}\;, \qquad\text{with}\quad d_j=2j+1\;.
\ee

The insertion of the generators in \eqref{insertion x1dotx2} can be done using
\beq \label{grasping}
\tau_i\otimes \tau^i = \sum_{m=-1}^{+1}\, (-1)^{1-m}\ L_m\otimes L_{-m}\;.
\ee
By plugging the above formulae into \eqref{insertion x1dotx2}, one gets the following modified intertwiner in terms of 3jm-symbols
\begin{multline}
\langle j_2, a_2';j_6,a_6';j_4,a_4\vert\,\mathcal{I}_{j_2 j_6 j_4}\vert 0\rangle \equiv \sum_{a_2, a_6,m} \begin{pmatrix} j_2 &j_6 &j_4\\a_2 &a_6 &a_4\end{pmatrix}\ \langle j_2,a_2'\vert\, L_m\,\vert j_2, a_2\rangle \ \langle j_6,a_6'\vert\, L_{-m}\,(-1)^{1-m}\ \vert j_6, a_6\rangle \\
= (-1)^{2j_2 +2j_6} N_{j_2}N_{j_6}\ \sum_{a_2, a_6,m} \begin{pmatrix} j_2 &j_6 &j_4\\a_2 &a_6 &a_4\end{pmatrix}\,(-1)^{j_2-a_2} \begin{pmatrix} 1 &j_2 &j_2\\ m &-a_2 &a_2'\end{pmatrix}\, (-1)^{1-m+j_6-a_6} \begin{pmatrix} 1 &j_6 &j_6\\ m &-a_6 &a_6'\end{pmatrix}\;.
\end{multline}
From the invariance under rotations on the indices $(a_2',a_6',a_4)$, it follows that this intertwiner must be proportional to the standard 3jm-symbol $\left(\begin{smallmatrix} j_2 &j_6 &j_4\\a_2' &a_6' &a_4\end{smallmatrix}\right)$. The proportionality coefficient is obtained by contracting $\mathcal{I}_{j_2 j_6 j_4}$ with this 3jm-symbol, and is found to be a Wigner 6j-symbol\footnote{The result comes via the following formula
\beq
\begin{Bmatrix} j_1 &j_2 &j_3\\ j_4 &j_5 &j_6\end{Bmatrix} = \sum (-1)^{\sum_{e=1}^6(j_e-a_e)} \begin{pmatrix} j_1 &j_2 &j_3\\a_1 &a_2 &a_3\end{pmatrix} \begin{pmatrix} j_3 &j_4 &j_5\\-a_3 &-a_4 &a_5\end{pmatrix} \begin{pmatrix} j_1 &j_5 &j_6\\-a_1 &-a_5 &a_6\end{pmatrix} \begin{pmatrix} j_2 &j_6 &j_4\\-a_2 &-a_6 &a_4\end{pmatrix}\;,
\ee
which is a standard definition of the 6j-symbol.
}, so that
\beq \label{x1dotx2}
\bigl(\widehat{\bigl(X_2\cdot X_6\bigr)}\ s_{\rm tet}^{\{j_e\}}\bigr)(g_1,\dotsc,g_6) = (-1)^{j_2+j_4+j_6}\,N_{j_2} N_{j_6}\,
\begin{Bmatrix} j_2 &j_2 &1\\j_6 &j_6 &j_4\end{Bmatrix}\ s_\Gamma^{\{j_e\}}(g_1,\dotsc,g_6)\;.
\ee
The fact that we find a spin 1 in the symbol is simply due to the fact that the indices contracted in the grasping \eqref{grasping} live on the vector, i.e. spin 1 representation.

The above expression will be very useful for us. Still, it can be simplified thanks to the known value of the above 6j-symbol, and observing that among the prefactors we find the eigenvalues of the length operators for $e_2, e_6$. This naturally leads to
\beq
\bigl(\widehat{\cos\phi_{e_2 e_6}}\ s_{\rm tet}^{\{j_e\}}\bigr)(g_1,\dotsc,g_6) = \frac{\bigl[j_2(j_2+1)+j_6(j_6+1) - j_4(j_4+1)\bigr]}{2\sqrt{j_2(j_2+1)\ j_6(j_6+1)}}\ s_\Gamma^{\{j_e\}}(g_1,\dotsc,g_6)\;.
\ee
This exactly reproduces the classical expression of the dihedral angles of a triangle, with  quantized lengths $\ell_e(j_e) = \sqrt{j_e(j_e+1)}$. Notice that a similar expression, with a prefactor going to 1 in the large length limit, was found though in a completely different setting based on discretizing the path integral on a triangulation of spacetime \cite{bf-aarc-val}.

At this stage, we are able to notice that the part $(X_2\cdot X_6)$ of $H_{v,f}$ produces a 6j-symbol and a phase factor which are exactly those entering the first term of the coefficient $A_0(j_1)$, \eqref{A_0}, which appears in the recurrence relation on the 6j-symbol.

\subsection{Quantization of the cosine of the 2d dihedral angle after parallel transport}

The next step is to compute the action of the same dot product after the insertion of the holonomy along the cycle. We first rewrite this quantity using the left derivatives $\tl{X}_e$,
\beq
X_2\cdot \Ad(g_2\mone g_1\mone g_6)(X_6) = \tl{X}_2\cdot\Ad(g_1\mone)(\tl{X}_6) = \sum_{i,j=x,y,z} \tl X_{2}^i\,\bigl( \delta_{ij} - R(g_1\mone)_{ij}\bigr)\,\tl X_{6}^j\;.
\ee
Then, at the quantum level, $\tl{X}^i_{2,6}$ will insert a generator $\tau^i$ on the left of the holonomies $g_{2,6}$, while $R(g_1\mone)$ will act as a multiplication. Let us focus on the part of the spin network function $s_{\rm tet}^{\{j_e\}}$ involving the matrix elements $\langle j_1, a_1\vert g_1\vert j_1, b_1\rangle$. Then, the operator $\tl{X}_2\cdot\Ad(g_1\mone)(\tl{X}_6)$ acting on
\beq \begin{split}
\ \sum_{\substack{a_1,b_1\\a_{2,3,5,6}}} (-1)^{j_1-a_1+j_6-a_6}\begin{pmatrix} j_1 &j_5 &j_6\\-a_1 &a_5 &-a_6\end{pmatrix} &\langle j_1, a_1\vert g_1\vert j_1, b_1\rangle
\begin{pmatrix} j_1 &j_2 &j_3\\ b_1 &-a_2 &-a_3\end{pmatrix} (-1)^{j_2-a_2+j_3-a_3} \\
&\vert j_5, a_5\rangle\otimes \langle j_3, a_3\vert \,\otimes\,
\langle j_6, a_6\vert g_6 \otimes \langle j_2, a_2\vert g_2\;,
\end{split}
\ee
gives the following expression
\beq \begin{split}
-\sum_{\substack{a_1,b_1, m, n\\a_{2,3,5,6}}} (-1)^{j_1-a_1+j_6-a_6}\begin{pmatrix} j_1 &j_5 &j_6\\-a_1 &a_5 &-a_6\end{pmatrix} &\langle 1,n\vert g_1\vert 1,m\rangle
\langle j_1, a_1\vert g_1\vert j_1, b_1\rangle
\begin{pmatrix} j_1 &j_2 &j_3\\ b_1 &-a_2 &-a_3\end{pmatrix} (-1)^{j_2-a_2+j_3-a_3} \\
\vert j_5, a_5\rangle &\otimes \langle j_3, a_3\vert \,\otimes\,
\langle j_6, a_6\vert L_{-n}(-1)^{1-n}\,g_6\, \otimes\, \langle j_2, a_2\vert L_{m} g_2\;.
\end{split}
\ee
Here we have re-expressed the insertions of generators in the basis $(L_m)$, which transforms with the Wigner matrix of spin 1, and we have used the fact that $(-1)^{m-n}\langle 1,-m\vert g_1\mone \vert 1,-n\rangle = \langle 1,n\vert g_1\vert 1, m\rangle$.

The product of two matrix elements of $g_1$ is expanded onto irreducible representations via
\begin{multline}
\langle 1,n\vert g_1\vert 1,m\rangle \langle j_1, a_1\vert g_1\vert j_1, b_1\rangle
= (-1)^{2j_1}\sum_{J_1=j_1-1}^{j_1+1}\sum_{A_1,B_1} d_{J_1}\,\langle J_1,A_1\vert g_1\vert J_1, B_1\rangle\\
\begin{pmatrix} 1 & j_1 & J_{1}\\ n & a_1 & -A_1\end{pmatrix}(-1)^{J_1-A_1}\,\begin{pmatrix} 1 & j_1 & J_1\\ -m & -b_1 & B_1\end{pmatrix}
(-1)^{1-m+j_1-b_1}\;,
\end{multline}
where $A_{1}=n+a_{1}$, $B_{1}=m+b_{1}$, while the action of the generators produces
\begin{align}
&\langle j_2, a_2\vert L_{m} = (-1)^{2j_2+1} N_{j_2}\sum_{a_2'} (-1)^{j_2-a_2'}\begin{pmatrix} 1 &j_2 &j_2\\m &-a_2' &a_2\end{pmatrix}\,\langle j_2, a_2'\vert\;,\\
&\langle j_6, a_6\vert (-1)^{1-n}L_{-n} = (-1)^{2j_6+1} N_{j_6}\sum_{a_6'} (-1)^{1-n+j_6-a_6'}\begin{pmatrix} 1 &j_6 &j_6\\-n &-a_6' &a_6\end{pmatrix}\,\langle j_6, a_6'\vert\;.
\end{align}
The contraction of the magnetic indices between the above formulae results in the addition of a link in the representation of spin 1 between the links carrying $J_1$ and $j_2$, and similarly between $J_1$ and $j_6$. It is then easy, using the same method as that used in the calculation of the dihedral angle, to extract 6j-symbols at the two vertices of $e_1$, in order to re-express the function in the spin network basis. After these manipulations, one gets
\beq \begin{split}
&\widehat{\bigl(X_2\cdot \Ad(g_2\mone g_1\mone g_6)(X_6)\bigr)}\ s^{\{j_e\}}_{\rm tet} \\
&= N_{j_2}N_{j_6} \sum_{J_1 = j_1-1}^{j_1+1} (-1)^{J_1+j_1+j_2+j_3+j_5+j_6+1}d_{J_1}\,\begin{Bmatrix} J_1 &j_1 &1 \\j_2 &j_2 &j_3\end{Bmatrix}\,\begin{Bmatrix} J_1 &j_1 &1 \\j_6 &j_6 &j_5\end{Bmatrix}\,s^{(J_1, j_2, \dotsc,j_6)}_{\rm tet}\;.
\end{split}
\ee
The prefactors, $N_{j_2}N_{j_6}$, are the same as those arising in the action of $X_2\cdot X_6$. In addition, one can recognize the coefficients $A_{\pm 1}(j_1)$, \eqref{Apm1}, in the terms $J_1=j_1\pm 1$, and the second part of the coefficient $A_0(j_1)$, \eqref{A_0}, for the term $J_1=j_1$. The technical point which requires to be careful is the sign of these coefficients. Finally, one obtains
\beq
\widehat{H}_{v,f}\ s^{\{j_e\}}_{\rm tet} = N_{j_2}N_{j_6}\biggl[
A_0(j_e)\,s^{\{j_e\}}_{\rm tet} + A_{-1}(j_e)\,s^{(j_1-1,j_2,\dotsc,j_6)}_{\rm tet} + A_{+1}(j_e)\,s^{(j_1+1,j_2,\dotsc,j_6)}_{\rm tet}\biggr]\;,
\ee
where the coefficients $ A_{m}(j)$ are given by the equation (\ref{A_0})

The last step is to derive the constraint equation \eqref{wdw eq} for a linear combination of spin network states
\beq
\psi(g_1,\dotsc,g_6) = \sum_{j_1,\dotsc,j_6}\Bigl[\prod_{e=1}^6 (2j_e+1)\Bigr] \psi(j_1,\dotsc,j_6)\,s^{\{j_e\}}_{\rm tet}(g_1,\dotsc,g_6)\;.
\ee
A brief examination of the coefficients $A_{\pm1}(j_1)$ reveals that: $d_{j_1\mp1}A_{\pm1}(j_1\mp1) = d_{j_1} A_{\mp1}(j_1)$. This way,
the equation $\widehat{H}_{v,f}\psi(g_1,\dotsc,g_6)=0$ exactly becomes the following difference equation of the second order
\be \label{wdweq}
A_{+1}(j_e)\,\psi(j_1+1,\dotsc,j_6) +
A_{0}(j_e)\,\psi(j_1,\dotsc,j_6) +
A_{-1}(j_e)\,\psi(j_1-1,\dotsc,j_6) = 0\;.
\ee
This is obviously the recurrence relation which defines the 6j-symbol from the Biedenharn-Elliott identity. We emphasize here that we have produced the exact equation, and not only a large spin approximation.

\subsection{Equivalence with flatness on trivalent plaquettes}

We now want to go further that a single tetrahedron, by showing how the Hamiltonian we propose enables to recover the projector onto the trivial holonomy sector for trivalent faces. This process can be thought of as a Pachner move 3-1 which reduces the triangulation by removing a dual face. Let us explain how our Hamiltonian implements it. In the dual picture to the triangulation, the initial configuration is a cycle bounding a face with three links. Thus we have three constraints $H_{v,f}$, that are good coordinates in the neighbourhood of $R(g_f)=\unit$. At the quantum level  we get three difference equations to impose the three constraints on the holonomy $g_f$ around the cycle.

We take the same notation as before, denoting $e_1, e_2, e_6$ the links along the cycle, and $e_3, e_4, e_5$ the links which connect to the cycle. This is like in Figure \ref{fig:tet}, except that the links $e_3, e_4, e_5$ do not necessarily meet. A key feature of the difference equation, which can be observed in the tetrahedral case, is that it only depends on the spins on the cycle and on those which connect to the cycle. It is independent of the fact that the latter ($e_3, e_4, e_5$ here) meet.

Using the three difference equations, we first want to show that the physical state factorizes on the cycle. In the spin network basis, the state is a function of the spins. In this configuration, its dependence on the spins $j_1,j_2,j_6$ actually completely factorizes,
\beq \label{factorization 6j}
\psi_{\rm phys}(j_e) = (-1)^{2j_3}\,\begin{Bmatrix} j_1 &j_2 &j_3\\j_4 &j_5 &j_6\end{Bmatrix}\ \phi(j_e')\;,
\ee
where the spins $j_e'$ are the spins of all links of the graph except $e_1, e_2, e_6$. This is known in the 3d gravity case from the projector onto physical states \cite{noui-perez-ps3d}, and work is in progress towards the precise loop quantization of higher dimensional BF theory \cite{discrete-bf}. Such a factorization property is actually very natural. Indeed, a flat connection induces on the boundary of a face with three links holonomies which are necessarily trivial up to gauge. Then, it is known that spin networks evaluated on the identity satisfy such factorizations, and the case of cycles with more links is also known \cite{recurrence-paper}.

The idea to extract the factorization is that the recursion relations hold on $j_1, j_2, j_6$. We can thus implement them from $\psi(j_1, j_2, j_6, j_e')$ so as to reach an initial state with $j_1=0, j_2 =j_3$ and $j_6 = j_5$. The factorization \eqref{factorization 6j} is then proved with the initial condition\footnote{It will be convenient in the following to explicitly factor a phase $(-1)^{2j_3}$. The basic reason is that in our conventions $e_3$ is oriented inwards the cycle while $e_4, e_5$ are outwards.}\footnote{Notice that $\left\{\begin{smallmatrix} 0 &j_3 &j_3 \\ j_4 &j_5 & j_5\end{smallmatrix}\right\} = (-1)^{j_3+j_4+j_5}/\sqrt{d_{j_3}d_{j_5}}$ is never zero.}
$\psi(j_e)_{|j_1=0} = (-1)^{2j_3} \left\{ \begin{smallmatrix} 0 &j_3 &j_3 \\ j_4 &j_5 & j_5\end{smallmatrix}\right\} \phi(j_e')$.

Now we are ready to present the consequences of the factorization \eqref{factorization 6j} in the group representation of wave functions, in which we also get a factorization of the cycle,
\beq
\psi_{\rm phys}(g_1,g_2,g_3,g_4,g_5,g_6,\dotsc) = \delta(g_2\mone\,g_1\mone\,g_6)\ \phi(g_2\mone g_3, g_4, g_5 g_6,\dotsc)\;,
\ee
with
\beq
\phi(g_2\mone g_3, g_4, g_5 g_6,\dotsc) = \sum_{\{j_e'\}} \Bigl[\prod_{e'} d_{j_e'}\Bigr]\ \phi(j_e')\ s^{\{j_e'\}}(g_2\mone g_3, g_4, g_5 g_6,\dotsc)\;.
\ee
We thus see that the wave function has support on holonomies with trivial parallel transport on the cycle, $g_2\mone\,g_1\mone\,g_6=\unit$. There is however one subtlety which is that relating two families of 6j-symbols whose spins differ by $\pm1/2$ for example has to be done by hand here. Thus, it may be convenient to restrict to the $\SO(3)$ model to avoid this. Otherwise, one has to relate the initial conditions of the different families through the spin 1/2 recursion \cite{varshalovich-book}.

To obtain the above equations, one writes the group Fourier transform of the physical topological state,
\begin{align}
\nonumber \psi_{\rm phys}(g_e) &= \sum_{\{j_e\}} \Bigl[\prod_{e} d_{j_e}\Bigr]\ \psi_{\rm phys}(j_e)\ s^{\{j_e\}}(g_e)\;,\\
&= \sum_{\{j_e'\}} \Bigl[\prod_{e'} d_{j_e'}\Bigr]\ \phi(j_e') \sum_{j_1,j_2,j_6} d_{j_1} d_{j_2} d_{j_6}\,\begin{Bmatrix} j_1 &j_2 &j_3\\j_4 &j_5 &j_6\end{Bmatrix}\ (-1)^{2j_3}\,s^{\{j_e\}}(g_e)\;.
\end{align}
Using the group averaging identity $\int dh\ D^{(j_1)}_{a_1 b_1}(h) D^{(j_2)}_{a_2 b_2}(h) D^{(j_3)}_{a_3 b_3}(h) = \Bigl(\begin{smallmatrix} j_1 &j_2 &j_3\\a_1 &a_2 &a_3\end{smallmatrix}\Bigr) \left(\begin{smallmatrix} j_1 &j_2 &j_3\\b_1 &b_2 &b_3\end{smallmatrix}\right)$, and the Fourier expansion of the Dirac delta on the group, $\delta(g) = \sum_j d_j \chi_j(g)$, the sums over $j_1, j_2, j_6$ can be explicitly performed,
\beq
\sum_{j_1,j_2,j_6} d_{j_1} d_{j_2} d_{j_6}\,\begin{Bmatrix} j_1 &j_2 &j_3\\j_4 &j_5 &j_6\end{Bmatrix}\ (-1)^{2j_3}\,s^{\{j_e\}}(g_e) = \delta(g_2\mone\,g_1\mone\,g_6)\ s^{\{j_e'\}}( g_2\mone g_3, g_4, g_5 g_6,\dotsc)\;.
\ee

To get the full projector onto the moduli space of flat connections, one needs the same result on all faces, which in general do not have necessarily three links. But it is possible to reduce the analysis to this case by performing a sequence of Pachner moves 2-2 to transform a given face with an arbitrary number of links to a face with three links. This is feasible since as well-known the 2-2 move is purely kinematical (it just relies on $\SU(2)$ invariance and not on the dynamics). We shall give a detailed account for this process elsewhere since in any case, it would still be necessary to check that the full projector satisfies all the constraints. This challenging issue is beyond the scope of the paper.

\section{Higher order equations} \label{sec:spin2}

When using holonomy variables, the quantum theory usually encounters an ambiguity corresponding to the choice of the $\SU(2)$ representation taken to evaluate these holonomies. Here, to obtain the above recurrence relation, we defined the operator $H_{v,f}$ using the flux variables as living in the vector representation, and the holonomy around the face is naturally in the spin $1$ representation. Then, this led to the evaluation of the Biedenharn-Elliott identity with a spin being $1$. But we could have chosen to express the dot products, like $X_2\cdot X_6$ in a representation of higher spin, so that the holonomy around the face is in a representation of integral spin, $J\in\N$. The question is then how it could affect the dynamical difference equation. A natural candidate for the resulting difference equation is the recurrence relation coming from restricting the Biedenharn-Elliott identity to spin $2$. The latter takes the following form:
\begin{multline} \label{spin2}
A_{+2}^{(2)}(j_1)\,\begin{Bmatrix} j_1+2 &j_2 &j_3\\j_4 &j_5 &j_6\end{Bmatrix} + A_{+1}^{(2)}(j_1)\,\begin{Bmatrix} j_1+1 &j_2 &j_3\\j_4 &j_5 &j_6\end{Bmatrix} + A_{0}^{(2)}(j_1)\,\begin{Bmatrix} j_1 &j_2 &j_3\\j_4 &j_5 &j_6\end{Bmatrix} \\
+ A_{-1}^{(2)}(j_1)\,\begin{Bmatrix} j_1-1 &j_2 &j_3\\j_4 &j_5 &j_6\end{Bmatrix} + A_{-2}^{(2)}(j_1)\,\begin{Bmatrix} j_1-2 &j_2 &j_3\\j_4 &j_5 &j_6\end{Bmatrix} = 0.
\end{multline}
The coefficients look like those of the previous relation, $A_{-1,0,+1}$, but for 6j-symbols with a spin 2,
\begin{align}
A_{\pm 2}^{(2)}(j_1) &= (-)^{2j_1+j_2+j_3+j_5+j_6+1}\bigl(2j_1+1\pm 4\bigr)\,\begin{Bmatrix} j_1\pm 2 &j_1 &2\\j_2 &j_2 &j_3\end{Bmatrix}\,\begin{Bmatrix} j_1\pm 2 &j_1 &2\\j_6 &j_6 &j_5\end{Bmatrix},\\
A_{\pm 1}^{(2)}(j_1) &= (-)^{2j_1+j_2+j_3+j_5+j_6}\bigl(2j_1+1\pm 2\bigr)\,\begin{Bmatrix} j_1\pm 1 &j_1 &2\\j_2 &j_2 &j_3\end{Bmatrix}\,\begin{Bmatrix} j_1\pm 1 &j_1 &2\\j_6 &j_6 &j_5\end{Bmatrix},\\
A_0^{(2)}(j_1) &= (-1)^{j_2+j_4+j_6}\begin{Bmatrix} j_2 &j_2 &2\\ j_6 &j_6 &j_4\end{Bmatrix} - (-)^{2j_1+j_2+j_3+j_5+j_6}\bigl(2j_1+1\bigr)\,\begin{Bmatrix} j_1 &j_1 &2\\j_2 &j_2 &j_3\end{Bmatrix}\,\begin{Bmatrix} j_1 &j_1 &2\\j_6 &j_6 &j_5\end{Bmatrix}.
\end{align}

In this section, we describe how to get this relation from the operator $H_{v,f}$, by taking it to higher spin. This means that we have to perform graspings on spin network nodes in the $\SU(2)$ representation of spin $2$. By this, we mean the elementary process which generates the 6j-symbol $\{\begin{smallmatrix} j_2 &j_2 &2\\ j_6 &j_6 &j_4\end{smallmatrix}\}$ appearing typically in $A_0^{(2)}$. As far as we know this technical point has not been discussed in the literature, so we here give some details.

The first natural idea is to square the dot product of the fluxes: $(\widehat{X_2\cdot X_6})^2$, but since at the quantum level the components $\widehat{X}_2^i, \widehat{X}_2^j$ do not commute, we take the symmetric part:
\beq
\what{X}_2^{(i}\,\what{X}_2^{j)} = \f12\bigl(\what{X}_2^i\,\what{X}_2^j + \what{X}_2^j\,\what{X}_2^i\bigr) = \what{X}_2^i\,\what{X}_2^j +\f12\,\bigl[\what{X}_2^j,\what{X}_2^i\bigr] = \what{X}_2^i\,\what{X}_2^j - \f{i}2\,\epsilon^{ij}_{\phantom{ij}k}\,\what{X}_2^k,
\ee
where in the last step we have used the quantum commutation relation $[\what{X}_2^j,\what{X}_2^i] = -i\epsilon^{ij}_{\phantom{ij}k}\,\what{X}_2^k$. We now use this symmetrized product to define the dot product of fluxes in the spin 2 representation:
\begin{align}
\bigl(\what{X_2 \cdot X_6}\bigr)^{(2)} &= \what{X}_2^{(i}\,\what{X}_2^{j)}\ \what{X}_{6(i}\,\what{X}_{6j)} - \f13 \what{X}_2^2\,\what{X}_6^2,\\
&= \bigl(\what{X_2 \cdot X_6}\bigr)^2 + \f12\,\what{X_2 \cdot X_6} - \f13 \what{X}_2^2\,\what{X}_6^2. \label{spin2grasping}
\end{align}
The reason for the last term, proportional to $\what{X}_2^2\,\what{X}_6^2$, is that it is necessary to produce the exact grasping for spin 2 as we will show. Indeed, from the previous results \eqref{length op}, \eqref{x1dotx2}, we know the action of $(\what{X_2 \cdot X_6})^{(2)}$ on a spin network state:
\beq
\widehat{\bigl(X_{2}\cdot X_{6}\bigr)}^{(2)}\ s_{\rm tet}^{\{j_e\}} = \biggl[N_{j_2}^2 N_{j_6}^2\,\begin{Bmatrix} j_2 &j_2 &1\\j_6 &j_6 &j_4\end{Bmatrix}^2 + \f{(-1)^{j_2+j_4+j_6}}{2}N_{j_2} N_{j_6}\,\begin{Bmatrix} j_2 &j_2 &1\\j_6 &j_6 &j_4\end{Bmatrix}
- \f{j_2(j_2+1)\,j_6(j_6+1)}{3} \biggr]\,s_{\rm tet}^{\{j_e\}}.
\ee
Parallelly, the Biedenharn-Elliott identity can be used to express the coefficient of the spin 2 grasping, that is: $\{\begin{smallmatrix} j_2 &j_2 &2\\j_6 &j_6 &j_4\end{smallmatrix}\}$, in terms of $\{\begin{smallmatrix} j_2 &j_2 &1\\j_6 &j_6 &j_4\end{smallmatrix}\}$ and $\{\begin{smallmatrix} j_2 &j_2 &0\\j_6 &j_6 &j_4\end{smallmatrix}\}$. The result is nothing but the above factor coming from the operator $\widehat{\bigl(X_{2}\cdot X_{6}\bigr)}^{(2)}$, so that the latter indeed produces the spin 2 grasping on spin network nodes:
\beq
\widehat{\bigl(X_{2}\cdot X_{6}\bigr)}^{(2)}\ s_{\rm tet}^{\{j_e\}} = \biggl[ \f16 (-1)^{j_2+j_4+j_6}N_{j_2}N_{j_6}\,\sqrt{d_{j_2-1}\,d_{j_2+1}\ d_{j_6-1}\,d_{j_6+1}}\,\begin{Bmatrix} j_2 &j_2 &2\\j_6 &j_6 &j_4\end{Bmatrix}\biggr]\,s_{\rm tet}^{\{j_e\}}.
\ee

Now that we know how to produce 6j-symbols with a spin 2, let us come back to our Hamiltonian operator and recurrence relations. In order to generate \eqref{spin2}, we propose the following operator:
\begin{align}
\what{H}_{v,f}^{(2)} &\,=\, \widehat{\bigl(X_{2}\cdot X_{6}\bigr)}^{(2)} \,-\, \widehat{\bigl(X_2\cdot \Ad(g_2\mone g_1g_6)(X_6)\bigr)}^{(2)},\\
&\,=\,  \bigl(\what{X_2\cdot X_6}\bigr)^2 \,-\, \Bigl(\what{X}_2\cdot \Ad(g_2\mone g_1 g_6)(\what{X}_6)\Bigr)^2 \,+\, \f12\,\what{X_2\cdot X_6} \,-\, \f12\,\what{X}_2\cdot \Ad(g_2\mone g_1 g_6)(\what{X}_6).
\end{align}
It turns out that this operator gives rise to the expected coefficients, up to an overall factor:
\beq
\what{H}_{v,f}^{(2)}\,s_{\rm tet}^{\{j_e\}} = \f16 (-1)^{j_2+j_4+j_6}N_{j_2}N_{j_6}\,\sqrt{d_{j_2-1}\,d_{j_2+1}\ d_{j_6-1}\,d_{j_6+1}} \sum_{k=-2}^{+2} A_{k}^{(2)}(j_1)\,s_{\rm tet}^{(j_1+k,j_2,\dotsc,j_6)}.
\ee
This can be checked the same way we derive the spin 2 grasping, that is by writing the coefficients $A_k^{(2)}(j_1)$ in terms of 6j-symbols with a spin 1, via the Biedenharn-Elliott identity, and compare with the action of $\what{H}_{v,f}^{(2)}$. Typically, the Biedenharn-Elliott identity gives:
\beq
\begin{Bmatrix} j_1+2 &j_1+1 &1\\j_2 &j_2 &j_3\end{Bmatrix} \begin{Bmatrix} j_1+1 &j_1 &1\\j_2 &j_2 &j_3\end{Bmatrix} = \f{\sqrt{d_{j_2-1}d_{j_2+1}}}{\sqrt{6\,d_{j_1+1}}\,N_{j_2}}\,\begin{Bmatrix} j_1+2 & j_1 &2\\j_2 &j_2 &j_3\end{Bmatrix}.
\ee
This can be lifted to a relation between the coefficients coming from the action of $H_{v,f}$ and those of \eqref{spin2}:
\beq
-A_{+1}(j_1+1)\,A_{+1}(j_1) =  \f{\sqrt{d_{j_2-1} d_{j_2+1}\,d_{j_6-1} d_{j_6+1}}}{6\,N_{j_2}\,N_{j_6}}\ A_{+2}^{(2)}(j_1).
\ee
Since the left hand side is the coefficient of $s_{\rm tet}^{(j_1+2,j_2,\dotsc,j_6)}$ when acting with $H_{v,f}^{(2)}$ on the spin network function, this shows the desired result for the highest shift on $j_1$. The other terms can be derived in the same way.

The final step is to write the constraint equation:
\beq
\what{H}_{v,f}^{(2)}\ \sum_{j_1,\dotsc,j_6}\Bigl[\prod_{e=1}^6 (2j_e+1)\Bigr] \psi(j_1,\dotsc,j_6)\,s_{\rm tet}^{\{j_e\}} = 0,
\ee
for an arbitrary expansion. By inspection of the coefficients, it comes that: $d_{j_1-k} A_{+k}^{(2)}(j_1-k) = d_{j_1} A_{-k}^{(2)}(j_1)$, so that we can write the following difference equation:
\beq
\sum_{k=-2}^{+2} A_k^{(2)}(j_1)\,\psi(j_1+k,j_2,\dotsc,j_6) \,=\,0.
\ee
However, in contrast with the difference equation coming from $H_{v,f}$, \eqref{wdweq}, the above constraint can not be solved with a single initial condition. Indeed, if one looks for instance at the minimum value taken by $j_1$, that is $j_1^{\min} = \max(\vert j_2-j_3\vert, \vert j_5-j_6\vert)$, the coefficients $A^{(2)}_{-2}(j_1^{\min}), A^{(2)}_{-1}(j_1^{\min})$ become zero. But the equation has three terms, and one has to provide two initial conditions: the values of $\psi(j_1^{\min})$ and $\psi(j_1^{\min}+1)$. So in this sense, the spin 2 difference equation is weaker than the spin 1 equation.

More generally, we can extract from the Biedenharn-Elliott identity a recurrence relation on the 6j-symbol of the form:
\beq
\sum_{k=-J}^{+J} A_k^{(J)}(j_1)\,\begin{Bmatrix} j_1+k &j_2 &j_3 \\j_4 &j_5 &j_6\end{Bmatrix} = 0,
\ee
for $J\in\N$. The coefficients $A^{(J)}_k(j_1)$ are products of 6j-symbols with a spin fixed to $J$, and we expect them to admit an expression in terms of 6j-symbols with a spin fixed to 1, through recursive use of the Biedenharn-Elliott identity. Typically, the highest coefficient should take the form
\beq
A^{(J)}_J(j_1) = (-)^{2j_1+j_2 +j_3 + j_5+j_6} d_{j_1+J} \begin{Bmatrix} j_1+J &j_1 &J\\j_2 &j_2 &j_3\end{Bmatrix}\begin{Bmatrix} j_1+J &j_1 &J\\j_6 &j_6 &j_5\end{Bmatrix} \propto \prod_{k=1}^J A_{+1}(j_1+k),
\ee
up to a normalization depending only on $j_2, j_6$. To determine the operator $H_{v,f}^{(J)}$ producing such an equation, one should first find the grasping operator in the representation of spin $J$. Finally, the corresponding Wheeler-DeWitt equation, $\what{H}_{v,f}^{(J)}\,\psi = 0$, will be a difference equation, solved by $J$ initial conditions.

\section{Conclusion}

In this paper, we have introduced a classical constraint $H_{v,f}$ on the phase space of loop quantum gravity on a single graph. This object naturally comes as a discretization of the scalar constraint $H=E^a_i E^b_j\epsilon^{ij}_{\phantom{ij}k} F_{ab}^k$, and is adapted to the specific smearing of the fields at the root of loop quantum gravity. In the holonomy-flux variables, it is the projection of the holonomy around a cycle of a spin network graph onto flux variables based on a vertex. From the point of view of the discrete geometry described in LQG, it amounts to saying that the holonomies enable to compute the dihedral angles between triangles as the dihedral angles of flat 3-simplices defined by the lengths. Moreover,
\begin{itemize}
 \item it is classically first class (at least on a specific simple triangulation of the sphere),
 \item the quantum equation $\hat{H}_{v,f}\vert \psi\rangle =0$ is realized in the spin network basis as a difference equation,
 \item on trivalent faces, the latter is a recurrence relation which is known to encode the symmetries at the quantum level in the spin foam formalism,
 \item and which is equivalent to projecting the face on its flat sector.
\end{itemize}
The facts that (in the case we study) the classical algebra is first class and that direct quantization of $H_{v,f}$ produces the correct and expected quantum constraints suggest some conclusions about the constraint algebra. The usual question of how to split and represent the diffeomorphism algebra in loop quantum gravity gets here a simple answer: it is not necessary to split it, but instead to project the curvature along all different normals to the triangles meeting at a vertex of the triangulation. Defining the discrete scalar constraint $H_{v,f}$ according to each of these normals is here sufficient.

To complete the analysis of three-dimensional gravity with those operators, some important things still remain to be done. Mainly, one needs to show that it is possible to recover the usual projector onto flat connections for surfaces of arbitrary genus, to find the structure functions of the algebra generated by the $H_{v,f}$ classically, and to check directly the quantum algebra. Preliminary calculations show that the projector is indeed annihilated by all the constraints. But getting an explicit quantum algebra is a difficult task.

Our result shows that it is possible to precisely identify the recurrence relations satisfied by spin foam amplitudes with a quantum implementation of the classical symmetries in loop quantum gravity. This suggests two ways to apply this programme to 4d LQG. The first idea is to derive recurrence relations for the amplitudes of the most promising spin foam models, and then try to produce them from an operator in LQG. A few recurrence relations for non-topological models were obtained in \cite{recurrence-paper}. Since these models describe geometry with areas and normals to triangles, we expect the corresponding operator to produce both differences on spins and some differential parts on the normals.

The second idea is to first derive difference equations from an operator and then interpret them as recurrence relations for some spin foam models. Typically, our operator $H_{v,f}$ has a natural generalization in 4d. There, the curvature tensor has more than one space component on the canonical hypersurface, so we may try to sum $H_{v,f}$ for three independent faces meeting at a vertex. For example, if $v$ is a 4-valent vertex, with edges labelled $1,2,3,4$, dual to a piece of triangulation, one may consider:
\beq \label{4dproposal}
H_{v,f_{12}} + H_{v,f_{23}} + H_{v,f_{31}} = 0.
\ee
Although certainly too naive, this can be interesting because, if $H_{v,f}$ is seen as the generator of the topological symmetry, then the above quantity will only imply invariance under a combination of the corresponding changes, as argued in \cite{recurrence-paper} on a toy example. Such an operator would generate a spin foam model, {\it a priori} different from those recently proposed.

Ultimately, one may expect these two approaches to coincide. This is actually what we achieved in the present article on the 3d model. In four dimensions, some preliminary results have been obtained in \cite{recursion-semiclass}. There the topological Ooguri model for BF theory is revisited by lifting the Hamiltonian we have just used here, $H_{v,f}$, to the boundary of a 4-simplex. Classically, this provides the phase space of 4d loop quantum gravity with a Hamiltonian dynamics which can be interpreted in terms of twisted geometries. In the spin network basis, the Wheeler-DeWitt equation gives recursion relations which are actually satisfied by the Wigner 15j-symbol. We expect to extend these ideas for more realistic spin foam models than the topological model.

More generally, the relation between spin foams and Hamiltonian dynamics is investigated in \cite{recursion-semiclass} in the large spin limit through difference equations on the 4-simplex amplitude. In 3d, such an equation is obviously the recursion relation \eqref{rec6j} on the 6j-symbol, simplified in the asymptotics. Such equations give criteria to know whether a model is semi-classically approximated by some quantum Regge calculus. For instance, it is clear from these results that the (naive) proposal \eqref{4dproposal} is solved in the asymptotics by exponentials of $i$ times the Regge action of the 4-simplex with quantized areas.

Thus, we think that difference and recurrence relations give a very efficient tool to: $i)$ build well-defined Hamiltonian operators and $ii)$ make the link with spin foam proposals. It would be interesting to use them to make contact and improve recent works which tackle the dynamics of quantum gravity. In particular, a new regularization of the Hamiltonian for general relativity has been proposed in \cite{alesci-reg}. A relation between the topological and some non-topological spin foam models and the canonical framework has been unravelled in \cite{alesci-ps}, but the physical meaning of the projection operators which are used should be clarified. The case of spin foams with finite groups has been recently introduced in \cite{baby-sf} as a testing ground to investigate such issues and also that of the continuum limit. 

That continuum limit issue was not a difficulty here because the theory is topological and the Hamiltonian needs not be graph changing, so that the continuum limit could be taken like in \cite{noui-perez-ps3d}. Nevertheless, the point remains challenging for four-dimensional gravity, for which recursion relations on a fixed graph can certainly not solve the model. Also in principle, since we have first quantize and then reduce the degrees of freedom at the quantum level, the issue of inhomogeneities which plague four-dimensional quantum gravity could be addressed in the context of the three-dimensional model, and it is not clear that our quantization provides any insight from that view.

In the framework of group field theory, a symmetry which generates a recursion relation (associated to the Pachner move 1-4) in the 3d colored model has been exhibited in \cite{diffeo-gft}. The recursion is a bit different from the one we have discussed here (which is in fact associated to the Pachner move 3-2), but the geometric interpretation is appealing. Moreover, it could be a way to find generic recursions on spin foam amplitudes for larger triangulations than a single simplex. A first step in that direction can be found in \cite{rec6jsquare}, still on a simple triangulation of the 3-sphere.

We want to emphasize that the relation between the Wheeler-DeWitt equation and the Biedenharn-Elliott is certainly not a coincidence. Actually the existence of a link between the set of flat connections on a surface and Wigner 6j-symbols is known for a while \cite{ooguri-3d, barrett-crane-wdw, freidel-louapre-PR1, noui-perez-ps3d} and more recently \cite{twisted-bubbles}, and our work rather sheds light on this intimate relationship. As we have seen in the section \ref{sec:quantum}, the recursion on the 6j-symbol and the structure of $\SU(2)$ recoupling are strong enough to project the holonomy around a triangular face onto the unit of the group.

Along this line, it may be possible to take advantage of the fact that the recursion relation and the BE identity for the homogeneously curved 3d spacetime is known from the quantum group $\SU(2)_q$ and the Turaev-Viro model, in order to find the corresponding operator in LQG. Important results and insights into this difficult issue have been recently obtained in \cite{noui-nonco-hol}, where the non-commutative connection which depends on the cosmological constant have been quantized. We think that it could be a good starting point to find the Hamiltonian operator which generates the recursion relations for the case of a non-zero cosmological constant.


\section*{Acknowledgements}

\noindent It is a pleasure to thank Bianca Dittrich for numerous discussions and for the incentive to carry out the project.

\noindent Research at Perimeter Institute is supported by the Government of Canada through Industry Canada and by the Province of Ontario through the Ministry of Research and Innovation.




\end{document}